\documentclass[fleqn,usenatbib]{mn2e}

\usepackage{aas_macros}
\usepackage{color}
\usepackage{bm}
\usepackage[colorlinks=true, allcolors=blue]{hyperref}
\usepackage{newtxtext,newtxmath}

\usepackage[T1]{fontenc}
\usepackage{ae, aecompl}
\usepackage{times}
\usepackage{amssymb}
\usepackage{float}
\usepackage{graphicx}
\usepackage{epstopdf}
\usepackage{lscape}
\usepackage{threeparttable}
\usepackage{bigints}
\usepackage{soul}
\usepackage[normalem]{ulem}

\newcommand{\kms}{km s$^{-1}$}
\newcommand{\Gaia}{{\it Gaia}}

\defcitealias{Marchetti+19}{M19}
\defcitealias{Marchetti21}{M21}

\title[{\Gaia} DR3 and hypervelocity stars]{{\Gaia} DR3 in 6D: The search for fast hypervelocity stars and constraints on the Galactic Centre environment}

\author[Marchetti, Evans \& Rossi]{
Tommaso Marchetti$^{1}$\thanks{E-mail:\href{mailto:tommaso.marchetti@eso.org}{tommaso.marchetti@eso.org}},
Fraser A. Evans$^{2}$,
Elena Maria Rossi$^{2}$\\
$^{1}$European Southern Observatory, Karl-Schwarzschild-Strasse 2, 85748 Garching bei M{\"u}nchen, Germany \\
$^{2}$ Leiden Observatory, Leiden University, PO Box 9513, NL-2300 RA Leiden, The Netherlands\\
}

\date{Accepted XXX. Received YYY; in original form ZZZ}

\begin{document}

\pagerange{\pageref{firstpage}--\pageref{lastpage}} \pubyear{2020}

\maketitle

\label{firstpage}

\begin{abstract}
The third data release (DR3) of the European Space Agency satellite {\Gaia} provides coordinates, parallaxes, proper motions, and radial velocities for a sample of $\sim 34$ million stars. We use the combined 6-dimensional phase space information to search for hypervelocity stars (HVSs), unbound stars accelerated by dynamical processes happening in the Galactic Centre. By looking at the kinematics of {\Gaia} DR3 stars in Galactocentric coordinates and by integrating their orbits in the Galactic potential, we do not identify any HVS candidates with a velocity higher than $700$ \kms and robustly observed kinematics. Assuming a scenario wherein the interaction between a stellar binary and the massive black hole Sgr A$^*$ is responsible for HVS ejections from the Galactic Centre, we derive degenerate limits on the ejection rate of HVSs and the slope of the initial mass function of the primary star among binaries in the Galactic Centre. Our results indicate that the HVS ejection rate is $\lesssim 8\times10^{-5}$ yr$^{-1}$ assuming a Salpeter mass function, and this upper limit becomes progressively smaller for an increasingly top-heavy mass distribution. A fiducial HVS ejection rate of $10^{-4}$ yr$^{-1}$ prefers a mass function slope $\lesssim -2.35$, disfavouring previously claimed top-heavy initial mass functions among stars in the Galactic Centre.
\end{abstract}


\section{Introduction}
\label{sec:intro}

The exquisite quality of the data produced by the European Space Agency (ESA) satellite {\Gaia} is revolutionizing our knowledge of the Milky Way, allowing us to resolve and characterize its stellar populations to an unprecedented level of detail \citep{gaia}. The third data release (DR3), out on June 13th, 2022 \citep{Vallenari+22, Babusiaux+22}, provides a massive amount of new information, including (but not limited to) $\sim 220$ million low-resolution spectra \citep{DeAngeli+22,Montegrifo+22}, astrophysical parameters for $\sim 470$ million stars \citep{Creevey+22}, $\sim 34$ million radial velocities \citep{Katz+22}, and catalogues for $10$ million variable stars \citep{Eyer+22} and $\sim 800 \ 000$ binaries \citep{Damerdji+22, Holl+22, Siopis+22}. The {\Gaia} DR3 catalogue spans $34$ months of observations, and, combined with the astrometry already provided by the early third {\Gaia} data release \citep[EDR3,][]{gaiaedr3} for $\sim 1.5$ billion stars\footnote{Since {\Gaia} EDR3 astrometry and photometry are part of {\Gaia} DR3 (they share the same list of sources), in this work we will always refer to {\Gaia} DR3 when discussing these quantities.}, is the largest stellar dataset ever produced, with possible applications ranging from Solar System objects \citep{Tanga+22}, to Galactic structure \citep{Drimmel+22, Recio-Blanco+22}, to extra-galactic sources \citep{Bailer-jones+22}. 

{\Gaia} DR3 radial velocities are obtained by the Radial Velocity Spectrometer (RVS) \citep{Katz+22, Blomme+22}, which takes spectra spanning a range of wavelengths between $847$ nm and $874$ nm near the Ca triplet \citep{Cropper+18}. {\Gaia} DR3 radial velocities are determined using two different methods depending on $G_\mathrm{RVS}$, the apparent magnitude of a star in the RVS band \citep{Sartoretti+22}. For bright stars ($G_\mathrm{RVS} \leq 12$), {\Gaia} DR3 radial velocities are computed as the median values of the individual epoch observations. For fainter stars ($12 < G_\mathrm{RVS} \lesssim 14$), radial velocities are instead the combinations of the epoch cross-correlation functions. The quoted radial velocity errors reflect the adopted method \citep{zucker03, Katz+22}.

The {\Gaia} DR3 subset with sky positions, parallaxes, proper motions, and radial velocities allows the full characterization of the kinematic and orbital properties for $\sim 34$ million stars, a factor of five higher than what was available as part of {\Gaia} EDR3 \citep{Katz+19,Seabroke+21}. This combined set of all-sky, homogeneous, and precise photometric/astrometric/spectroscopic measurements is the ideal dataset in which to search for new hypervelocity stars (HVSs), stars coming from the Galactic Centre (GC) with velocities larger than the Galactic escape speed \citep[see][]{brown15}. The fastest HVSs can reach velocities higher than $\sim1000$ \kms and fly away from the GC on almost radial trajectories. Following the first serendipitous detection of an HVS candidate \citep{brown+05}, several works tried to identify new HVSs using a combination of photometric, astrometric and spectroscopic techniques \citep[e.g.][]{Brown+06, Brown+09, kollmeier+09, kollmeier+10, pereira+12, palladino+14, hawkins+15, marchetti+17, Marchetti+19, hattori+18b,  Bromley+18, Luna+19, Li+21, Marchetti21, Prudil+22}, but the majority of candidates have been shown to be either bound to the Galaxy or non-consistent with coming from the GC \citep{Boubert+18}. Even if a few very promising HVS candidates have been discovered in the outer halo of the Galaxy with blue colours and high velocities \citep[e.g.][]{brown+15, Brown+18}, large uncertainties in proper motions and distances have prevented a clear identification of their ejection location -- an origin in the stellar disc remains a possibility for many sources \citep[e.g.][]{irrgang+18, Kreuzer+20}. The only detected HVS with an unambiguous trajectory pointing directly radially away from the Galactic Centre is S5-HVS1, an A-type star at a distance of $\sim 9$ kpc from the Sun with a total velocity of $\sim 1700$ {\kms} in the Galactocentric rest frame \citep{Koposov+20}.

The most promising mechanism to explain the unbound velocities of HVSs involves the three body dynamical interaction between a stellar binary and the massive black hole at the centre of the Milky Way \citep[the Hills mechanism;][]{hills88}. A variation on this scenario involves the ejection of single stars following the interaction with a binary massive black hole, assuming the existence of an intermediate-mass black hole companion orbiting around Sgr A$^*$ \citep[e.g.][]{yu&tremaine03, Levin06, sesana+06, sesana+09, Rasskazov+19}. While other mechanisms have been suggested \citep[e.g.][]{oleary+08, abadi+09, capuzzodolcetta+15, Evans+20}, the presence of a massive black hole (and therefore an origin in the GC) is necessary to explain the extremely high velocities of HVS1 \citep{brown+05} and S5-HVS1 \citep{Koposov+20}. Another possible source of HVSs is the Large Magellanic Cloud (LMC), the most massive satellite galaxy of the Milky Way \citep[e.g.][]{boubert+16, Erkal+19, Evans+21}.

Thanks to their extreme velocities and the large range of distances spanned by their trajectories, HVSs have been proposed as dynamical probes to constrain the gravitational potential of the Milky Way \citep[e.g.][]{gnedin+05, yu&madau07, Contigiani+19, Gallo+21}, and even to test modified theories of gravity \citep{Chakrabarty+22}. Furthermore, their properties can be used to infer properties of the GC when adopting a particular ejection mechanism \citep[e.g.][]{Rossi+17, Evans+22, Evans+22b}. Specifically, \citet{Evans+22, Evans+22b}, assuming the Hills mechanism, showed that the population of HVSs predicted to be included in the {\Gaia} radial velocity  catalogue(s) strongly depends on the HVS ejection rate $\eta$ and on the slope $\kappa$ of the initial mass function for the primary star in GC binaries, $dN/dm \propto m^\kappa$. The absence of HVSs with radial velocities and precise parallaxes in {\Gaia} DR2 and EDR3 \citep[see][]{Marchetti+19, Marchetti21}, together with the detection of S5-HVS1 in the S$^5$ survey \citep{Li+19, Koposov+20}, constrains $\eta \sim 10^{-4}$ yr$^{-1}$ \citep{Evans+22b} for a top-heavy mass function in the GC \citep{lu+13}. The two parameters are strongly degenerate, with $\eta \sim 10^{-3}$ yr$^{-1}$ being preferred for a canonical \cite{salpeter55} mass function.

While a conspicuous population of HVSs is expected to be \textit{in principle} observed by the {\Gaia} satellite, the bulk of the stars are expected to be too faint to get a validated radial velocity measurement, making their identification nontrivial \citep{marchetti+18, Evans+21}. Still, the discovery of even a single HVS, or a clear non-detection of HVSs in {\Gaia} DR3, can be directly converted into updated constraints on the stellar population and dynamical processes happening in the GC, once the {\Gaia} selection function \citep[e.g.][]{Boubert+20I, Boubert+20II, Rix+21, Rybizki+21, Everall+22} is modelled. In this work, we report our search for HVSs from among the $\sim 34$ million stars in {\Gaia} DR3 with full phase space information.

This paper is organized as follows. In Section \ref{sec:method} we discuss the method we use to compute distances and velocities for all stars in {\Gaia} DR3 with a radial velocity measurement. Then, in Section \ref{sec:results}, we explain how we select the most promising HVS candidates, characterize their orbital properties, and discuss them individually. In Section \ref{sec:constr} we model our findings to extract constraints on the HVSs ejection mechanism and GC stellar population. Finally, we summarize and discuss our results in Section \ref{sec:discussions}.

\section{Method}
\label{sec:method}

{\Gaia} DR3 provides radial velocities ($v_\mathrm{rad}$) in the Solar System barycentric reference frame for a sample of $\sim 34$ million stars with magnitudes in the RVS band $G_\mathrm{RVS} \leq 14$ and effective temperatures in the range $3100 \ \mathrm{K} \leq T_\mathrm{eff} \leq 14500$ K \citep{Katz+22, Blomme+22}. Mean RVS spectra are available for $999 645$ stars \citep{Recio-Blanco+22b, Seabroke+22}. A subset of $33 653 049$ of the sources with radial velocities has also a full {\Gaia} astrometric solution: sky positions $(\alpha, \delta)$, parallaxes $\varpi$ and proper motions $(\mu_{\alpha *} \equiv \mu_\alpha \cos\delta, \mu_\delta)$, with corresponding uncertainties and correlations \citep{Lindegren+21a}. This combined set of stars with full phase space information can be used to derive three-dimensional positions/velocities and to reconstruct Galactic orbits. To do so, we start by correcting all {\Gaia} DR3 parallaxes by the parallax zero point $\varpi_\mathrm{zp}$, using the approach described in \citet{Lindegren+21b}. We also correct {\Gaia} DR3 radial velocities for cool ($T_\mathrm{eff} < 8500$ K) and hot ($T_\mathrm{eff} \geq 8500$ K) stars using the approach described, respectively, in \citet{Katz+22} and \citet{Blomme+22}. 

In this work, we choose to restrict our search for HVSs to the sample of stars with precise parallaxes, which can be simply inverted to derive an accurate geometric distance: 
\begin{equation}
    \label{eq:par1}
    \varpi - \varpi_\mathrm{zp} > 0 \ ,
\end{equation}
\begin{equation}
    \label{eq:par2}
    \frac{\sigma_\varpi}{\varpi - \varpi_\mathrm{zp}} < 0.2 \ ,
\end{equation}
where $\sigma_\varpi$ is the {\Gaia} DR3 parallax uncertainty. Deriving distances for fainter stars with larger relative errors in parallax can be done implementing a Bayesian approach \citep[see e.g.][]{bailer-jones, astraatmadjaI, bailer-jones+18, Luri+18, bailer-jones+21}, but this requires specifying prior knowledge on the distribution of stars in the Galaxy. In this work, we decide to adopt a conservative approach and not bias our results on the choice of the Galactic prior, focusing solely on the subset of $31129130$ stars ($\sim 92$\%) with precise and positive parallaxes.

Following the approach outlined in \citet{Marchetti+19} and \citet{Marchetti21}, we implement a Monte Carlo (MC) scheme to sample the distributions of distances and velocities given the quoted {\Gaia} observables and their uncertainties. We assume that the astrometry and the radial velocities are not correlated and we draw $100$ MC samples for each star from a four-dimensional multivariate Gaussian distribution centred on the quoted values of the observables $(\varpi, \mu_{\alpha*}, \mu_\delta, v_\mathrm{rad})$, with a covariance matrix constructed using the quoted {\Gaia} DR3 astrometric uncertainties and correlations. We assume that sky position uncertainties are negligible. 

The {\Gaia} DR3 astrometric processing pipeline to derive the astrometric solution uses {\Gaia} DR2 $G_\mathrm{BP} - G_\mathrm{RP}$ colours to calibrate the point spread function. When the photometric colour was not available, it was estimated from the astrometric solution (the \emph{astrometric pseudocolour}) through the chromatic displacement of the centroid of the images \citep{Lindegren+21a}. Therefore, for a subset of all sources in {\Gaia} DR3 ($\sim 66\%$), the astrometric model was used to fit 6 parameters instead of 5, and the resulting determination of parallaxes and proper motions is intrinsically less accurate \citep{Lindegren+21a}. We find that $766802$ ($2.5\%$) of the stars with radial velocities and full astrometry have a six-parameters solution. For this subset of sources, we construct a five-dimensional covariance matrix using the provided correlations between all the fitted astrometric parameters, including the \textsc{pseudocolour} \citep[see Section 6 in][]{Marchetti21}, which we use to correct each MC parallax sample for its corresponding zero-point offset.

To derive distances and total velocities, we assume a distance of the Sun to the Galactic Centre of $8.122$ kpc \citep{Gravity+18} and a height of the Sun above the Galactic disc of $z_\odot = 20.8$ pc \citep{Bennet+19}. Furthermore, we fix the three-dimensional Solar velocity to $\mathbf{v_\odot} = [12.9, 245.6, 7.78]$ \kms \citep{Reid+04, drimmel+18}. We then determine distances and total velocities in the Galactocentric rest-frame for all the stars, which we will use in the next sections to identify the most promising HVS candidates in {\Gaia} DR3. The complete characterization of the population of high velocity stars in {\Gaia} DR3 will be the focus of a companion paper.

\section{Searching for HVS candidates}
\label{sec:results}

In this Section, we use the catalogue of distances and total velocities derived in Section \ref{sec:method} to search for unbound HVSs with orbits consistent with an ejection from the centre of the Galaxy.

\subsection{Selecting robustly measured stars}
\label{sec:clean}

When searching for kinematic outliers, it is essential to select stars with accurate measurements, since an erroneous parallax, proper motion or radial velocity determination can translate into an artificially large total velocity \citep[see e.g.][]{Boubert+19, Seabroke+21, Katz+22}. To this end, we select stars in the {\Gaia} DR3 catalogue that satisfy the following quality criteria (see also Table ~\ref{tab:cuts}):
\begin{equation}
    \label{eq:ruwe}
    \textsc{ruwe} < 1.4 \ ,
\end{equation}
\begin{equation}
    \label{eq:rv_nb_transits}
    \textsc{rv\_nb\_transits} \geq 10 \ ,
\end{equation}
\begin{equation}
    \label{eq:cut_SNR}
    \textsc{rv\_expected\_sig\_to\_noise} \geq 5  \ .
\end{equation}
Above, \textsc{ruwe} is the Renormalised Unit Weight Error, corresponding to the magnitude and colour-renormalized square root of the reduced chi-squared statistic to the astrometric fit \citep{Lindegren+21a}. A large value of $\textsc{ruwe}$ could be due to unresolved multiplicity \citep[e.g.][]{Belokurov+20, Penoyre+20, Penoyre+22}, and in general it indicates a poor determination of the star's parallax and proper motion. \textsc{rv\_nb\_transits} is the number of epochs used to determine the radial velocity of each star, and a high value is recommended to select stars with accurate measurements. Finally, \textsc{rv\_expected\_sig\_to\_noise} is the expected signal-to-noise ratio in the combination of individual spectra used to determine the {\Gaia} DR3 radial velocity. \citet{Katz+22} show that a cut like equation \eqref{eq:cut_SNR} is the most efficient to remove spurious high radial velocities ($ > 750$ \kms) which might contaminate our search for HVSs. 

The quality cuts used in this work are summarised in Table \ref{tab:cuts}, where we see that a total of $\sim 20$ million stars satisfy the cuts in equations \eqref{eq:ruwe} to \eqref{eq:cut_SNR}.

\begin{table}
\caption{Summary of the cuts employed to select a clean subset of HVS candidates. $N_\mathrm{stars}$ denotes the number of stars in {\Gaia} DR3 surviving after each selection cut. The cuts are discussed in Sections \ref{sec:method}, \ref{sec:clean}, and \ref{sec:results:identification}.}
\label{tab:cuts}      
\centering                          
\begin{tabular}{l r}        
\hline               
Cut & $N_\mathrm{stars}$ \\    
\hline
   stars in {\Gaia} DR3 & 1 811 709 771 \\
   radial velocity from {\Gaia} DR3  & 33 812 183 \\
   full astrometric solution from {\Gaia} DR3 & 33 653 049 \\
   positive and precise parallaxes (eq. \ref{eq:par1} and \ref{eq:par2}) & 31 129 130 \\
   \textsc{ruwe} $<1.4$ & 27 901 700 \\      
   \textsc{rv\_nb\_transits} $\geq 10$ & 24 662 999\\
   \textsc{rv\_expected\_sig\_to\_noise} $\geq 5$ & 19 765 578 \\
   $v_\mathrm{GC} > 500$ \kms & 414 \\
   $v_\mathrm{r_{GC}} > 500$ \kms, $v_\mathrm{tan_{GC}} < 100$ \kms & 3 \\
   $r_\mathrm{min} < 1$ kpc & 2 \\
\hline                                   
\end{tabular}
\end{table}

\subsection{Identification of HVS candidates}
\label{sec:results:identification}

\begin{figure*}
	\centering
 	\includegraphics[width=0.75\textwidth]{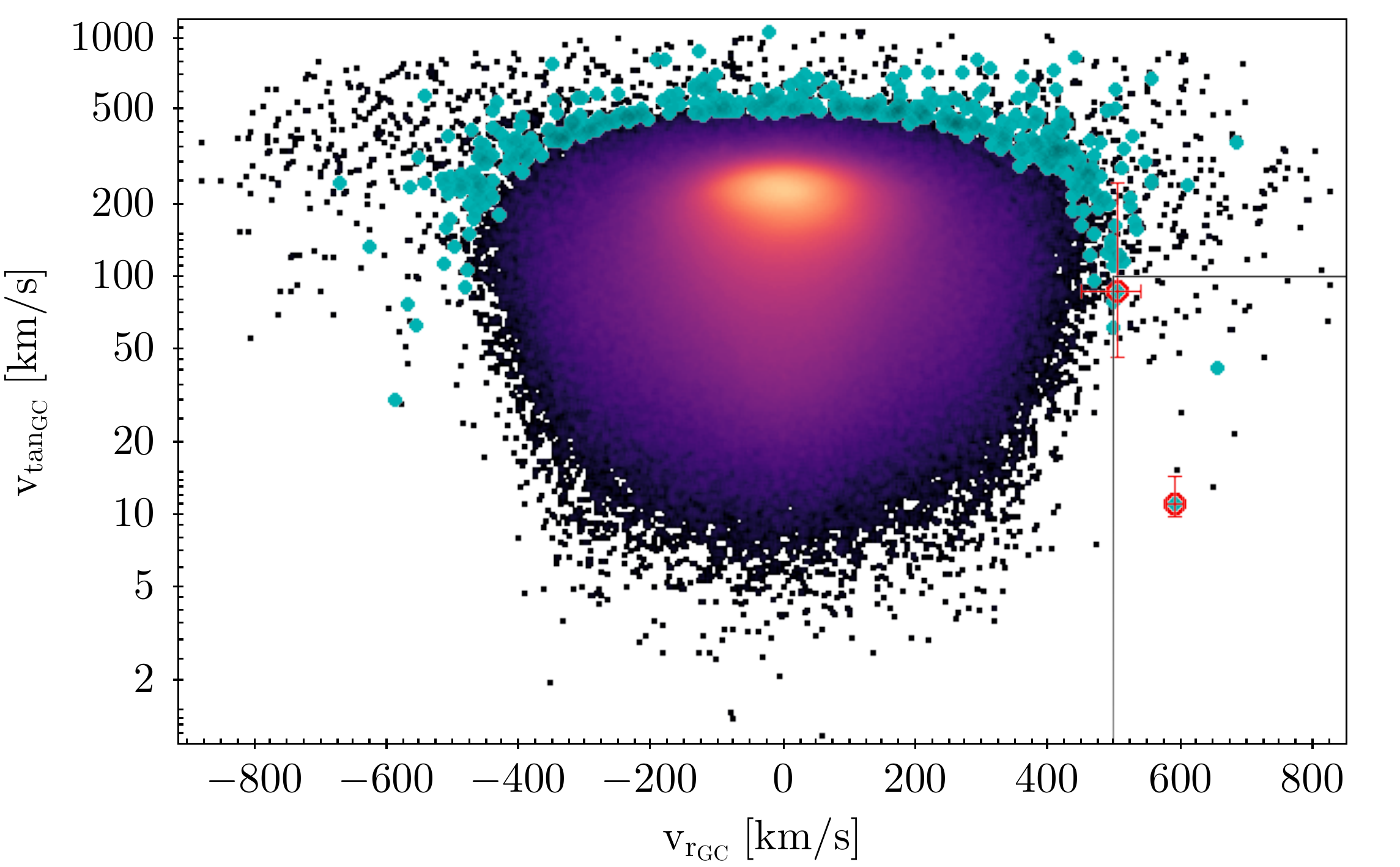}
	\caption{Galactocentric tangential velocity as a function of Galactocentric radial velocity for the $31129130$ stars satisfying the parallax cuts (see Section~\ref{sec:method}). The cyan points correspond to the clean sample of $414$ stars with $v_\mathrm{GC} > 500$ \kms (see Section \ref{sec:results:identification}). The grey lines mark the selection box $v_\mathrm{r_{GC}} > 500$ \kms, $v_\mathrm{tan_{GC}} < 100$ \kms. The red points correspond to the 2 HVS candidates we discuss in Section \ref{sec:HVSs}.}
	\label{fig:vR_vtan}
\end{figure*}

\begin{figure}
	\centering
 	\includegraphics[width=\columnwidth]{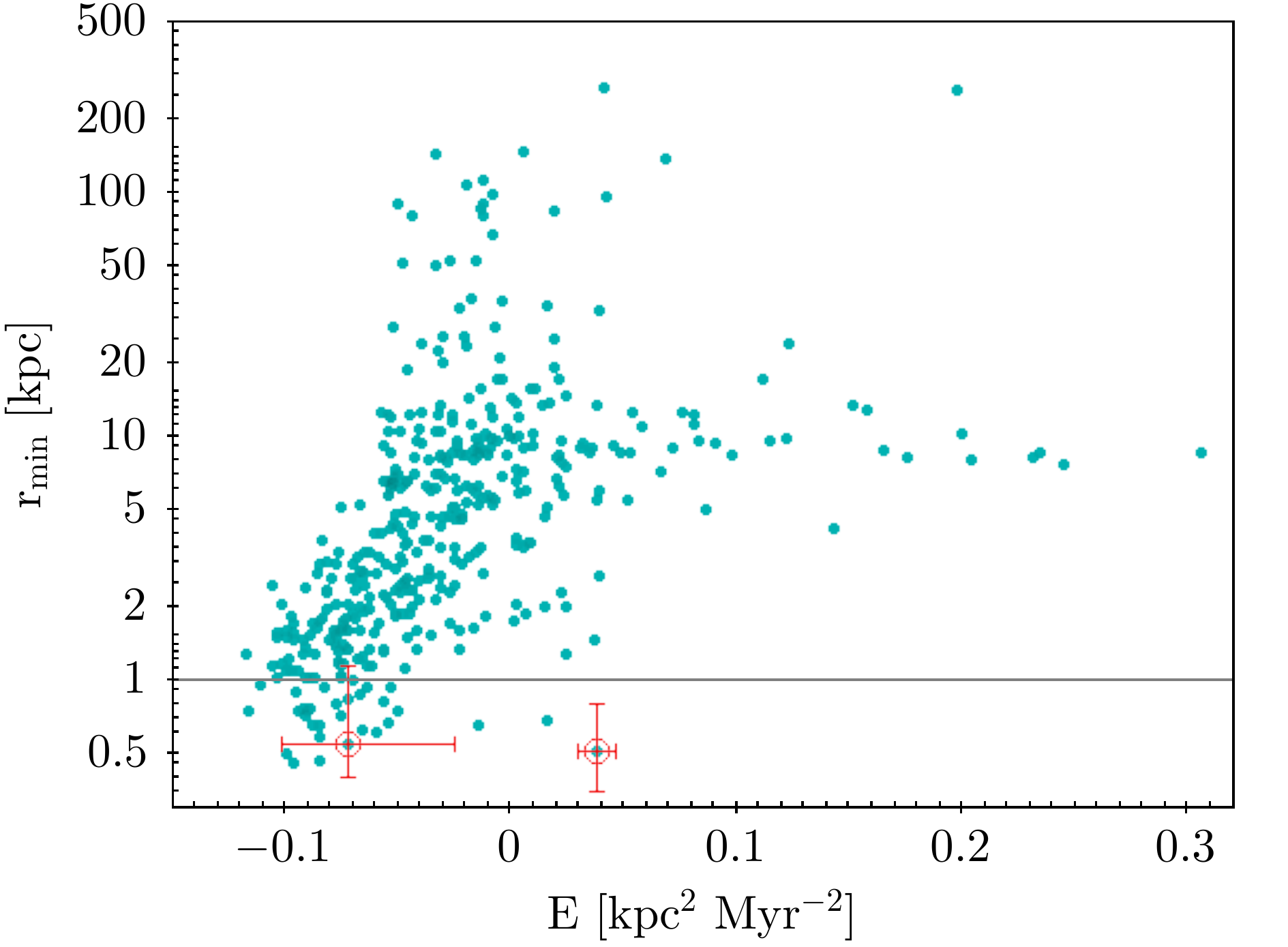}
	\caption{Minimum crossing radius $r_\mathrm{min}$ as a function of orbital energy $E$ for the subset of $414$ clean stars with $v_\mathrm{GC} > 500$ \kms. The horizontal line denotes the cut $r_\mathrm{min} < 1$ kpc, used to search for objects consistent with coming from the GC. The two candidates discussed in Section \ref{sec:HVSs} are shown with red open circles.}
	\label{fig:E_rmin}
\end{figure}

In our search for the fastest stars in the Galaxy, we restrict our sample to the stars with median total velocities in the Galactocentric rest-frame\footnote{Compared to \citet{Marchetti21}, in this work we decide not to use the probability to be unbound from the Galaxy $P_\mathrm{ub}$ to select high velocity stars, which requires a specific choice for the Galactic potential. We focus instead on high velocities to make the minimum number of assumptions in our search for HVSs.} $v_\mathrm{GC} > 500$ \kms, of the order of escape speed from the Sun's location \citep[e.g.][]{williams+17}. A total of $414$ stars satisfy this cut and will be the main focus of the following Sections.

To identify possible HVS candidates unbound to the Galaxy consistent with coming from the Galactic Centre, in Fig.~\ref{fig:vR_vtan} we show the distribution of Galactocentric tangential velocities $v_\mathrm{tan_{GC}}$ as a function of Galactocentric radial velocities $v_\mathrm{r_{GC}}$. We can clearly see that the majority of the stars are co-rotating with the Sun in the stellar disc, at $v_\mathrm{r_{GC}} \sim 0$, $v_\mathrm{tan_{GC}} \sim 230$ \kms. A conspicuous population of stars with a large spread in Galactocentric radial velocities, likely associated to the {\Gaia}-Sausage-Enceladus merger event \citep{Belokurov+18, Helmi+18}, is visible as well at lower Galactocentric tangential velocities. Finally, we expect unbound HVSs ejected from the GC to have large positive values of $v_\mathrm{r_{GC}}$ and low values of $v_\mathrm{tan_{GC}}$, appearing therefore in the bottom-right corner of Fig.~\ref{fig:vR_vtan}. This is because their trajectory from the GC through the halo is negligibly deflected from a radial orbit by asymmetry in the bulge, the stellar disc, triaxiality of the Galactic dark matter halo mass distribution, and/or the presence of the LMC \citep{Kenyon+18, boubert+20LMC}. To select the most promising HVSs, we thus select stars with median $v_\mathrm{r_{GC}} > 500$ \kms and $v_\mathrm{tan_{GC}} < 100$ \kms. A total of 3 stars satisfies these cuts.

For the $414$ stars with high Galactocentric total velocities ($v_\mathrm{GC} > 500$ \kms), we perform orbital integration to characterize their orbits and identify their possible ejection location. We use the \textsc{python} package \textsc{gala} \citep{gala} to integrate backwards in time the orbit of each star for a total time of $1$ Gyr with a time step of $0.1$ Myr. We adopt the default \textsc{gala} potential \textsc{MilkyWayPotential} \citep{bovy15}, an axisymmetric potential which is the sum of four Galactic components \citep[see also][for the adopted parameters]{Marchetti21}. The escape velocity to infinity from this chosen potential is $\sim 560$ \kms at the Solar position, consistent with estimates from \citet{smith+07, kafle+14, williams+17, monari+18, Prudil+22}. Following the method described in Section \ref{sec:method}, we draw $5000$ MC realizations of each star's observed phase space information and integrate each orbit backwards in time in the Galactic potential. For each orbit, we compute the eccentricity $e$, the energy $E$, the angular momentum $L_{\rm z}$, and the maximum vertical distance from the Galactic plane $|Z_\mathrm{max}|$.

In our search for stars coming from the GC, we keep track of each disc crossing (Galactic latitude $b = 0$). In the case of unbound ($E > 0$) orbits, there is only one disc crossing, however, for bound MC realizations ($E < 0$), there might be multiple crossings of the disc during the past Gyr. We then define the crossing radius $r_{\rm c}$ as the distance of the star from the GC at each disc crossing. To test the consistency of an HVS candidate with originating from the GC, we define the minimum crossing radius $r_\mathrm{min}$ as the minimum value of $r_{\rm c}$ attained during the $1$ Gyr orbital integration. For each orbit, we also compute the ejection velocity $v_\mathrm{ej}$ (defined as the total velocity in the Galactocentric rest frame at $r_\mathrm{min}$), and the flight time $t_\mathrm{F}$ (the time passed since $r_\mathrm{min}$).

Fig. \ref{fig:E_rmin} shows $r_\mathrm{min}$ as a function of orbital energy for the sample of $414$ stars back-propagated in the Galactic potential. The fastest HVSs are expected to reside in the bottom right of this plot, having large positive values of $E$ (unbound trajectories) and low values of $r_\mathrm{min}$ (consistent with coming from the GC). We find that $2$ stars out of the $3$ with $v_\mathrm{r_{GC}} > 500$ \kms and $v_\mathrm{tan_{GC}} < 100$ have $r_\mathrm{min} < 1$ kpc, and will be the main focus of the next subsection.

\subsection{HVS candidates}
\label{sec:HVSs}

In this Section, we describe individually the most promising candidates found in our systematic search for HVSs in the {\Gaia} DR3 catalogue: the $2$ stars with clean measurements, $v_\mathrm{r_{GC}} > 500$ \kms, $v_\mathrm{tan_{GC}} < 100$ \kms, and $r_\mathrm{min} < 1$ kpc. These stars are shown as red circles in Fig. \ref{fig:vR_vtan} and Fig. \ref{fig:E_rmin}.

\begin{itemize}

    \item {\Gaia} DR3 3126801097033888768 $(\alpha = 100.863669^\circ$, $\delta = 2.140983^\circ)$ is the most promising HVS candidate. For this star, we determine $v_\mathrm{r_{GC}} \ = 591 \pm 13$ \kms, $v_\mathrm{tan_{GC}} \ = 11^{+4}_{-1}$ \kms, $v_\mathrm{GC} \sim 600$ \kms. The {\Gaia} DR3 processing of the low resolution spectra using the A library of synthetic stellar spectra results in an effective temperature $T_\mathrm{eff} = 14362 \pm 125$ K, a surface gravity $\log g = 3.80 \pm 0.02$, and a metallicity of $\sim -1$. Even if the orbital integration predicts $r_\mathrm{min} = 505^{+284}_{-156}$ pc, further caution is needed to interpret its origin. The  star is observed at a Galactic latitude $b \sim -0.8^\circ$, corresponding to a Cartesian distance of $z \sim -24$ pc below the Galactic plane, and its orbit runs parallel to the Galactic disc in the direction of the GC. The most recent disc crossing occurred 2 Myr ago at a distance of $\sim 10$ kpc from the GC. After travelling in the plane of the disc ($|z| < 100$ pc) for $\sim 15$ Myr, the star crosses again the plane at $\sim 500$ pc from the GC, which results into the low value of $r_\mathrm{min}$ shown in Fig. \ref{fig:E_rmin}. The association of {\Gaia} DR3 3126...8768 with a GC origin is therefore poorly constrained, and this is confirmed also when adopting different models for the Galactic potentials \citep{Law+10, bovy15}, and when using the sampling of the {\Gaia} DR3 distance posterior (\textsc{distance\_gspphot} $= 4159^{+191}_{-174}$ pc).
    
    \item {\Gaia} DR3 6023361538639059840 $(\alpha = 243.605409^\circ$, $\delta = -34.296673^\circ)$ is the second object falling in the selection boxes used in this work, and its measurements are significantly less precise than those for {\Gaia} DR3 3126...8768. This is a red giant star with $v_\mathrm{r_{GC}} \ = 506^{+34}_{-55}$ \kms, $v_\mathrm{tan_{GC}} \ = 86^{+160}_{-41}$ \kms, $v_\mathrm{GC} \sim 600$ \kms. It is observed at $b \sim +12^\circ$, and its trajectory crosses the Galactic disk at $r_\mathrm{min} \sim 600$ pc from the GC, with a velocity $v_\mathrm{ej} \sim 630$ \kms, suggesting a possible origin in the GC. {\Gaia} DR3 6023...9840 is bound to the Galaxy in all the potential models explored in this work: it has a distance from the GC of $\sim 3$ kpc, where the escape speed from the Galaxy is higher than $600$ \kms \citep[e.g.][]{williams+17, monari+18}. Given the multiple disk crossings during its orbital integration, a clear identification of the GC as the origin for {\Gaia} R3 6023...9840 is not possible (we refer the reader to the discussion in Section \ref{sec:discussions} on \emph{bound} HVSs). We note that {\Gaia} DR3 astrophysical parameters are not available for this source.
    
\end{itemize}

Given the highly uncertain nature of the candidates discussed above, we decide to focus the model comparison in Section \ref{sec:constr} to stars with velocities above $700$ \kms. To search for possible HVSs, we also inspect individually all the stars with $v_\mathrm{GC} > 600$ \kms, relaxing the cuts in equations \eqref{eq:rv_nb_transits} and \eqref{eq:cut_SNR}, and find no unbound HVS candidate. Therefore, we conclude that we find no evidence for high-confidence HVSs in {\Gaia} DR3 with Galactocentric total velocity $v_\mathrm{GC} > 700$ \kms satisfying the selection cuts in equations \eqref{eq:par1} to \eqref{eq:cut_SNR}.

\section{Constraints on the Galactic Centre population}
\label{sec:constr}

\begin{figure}
	\centering
 	\includegraphics[width=\columnwidth]{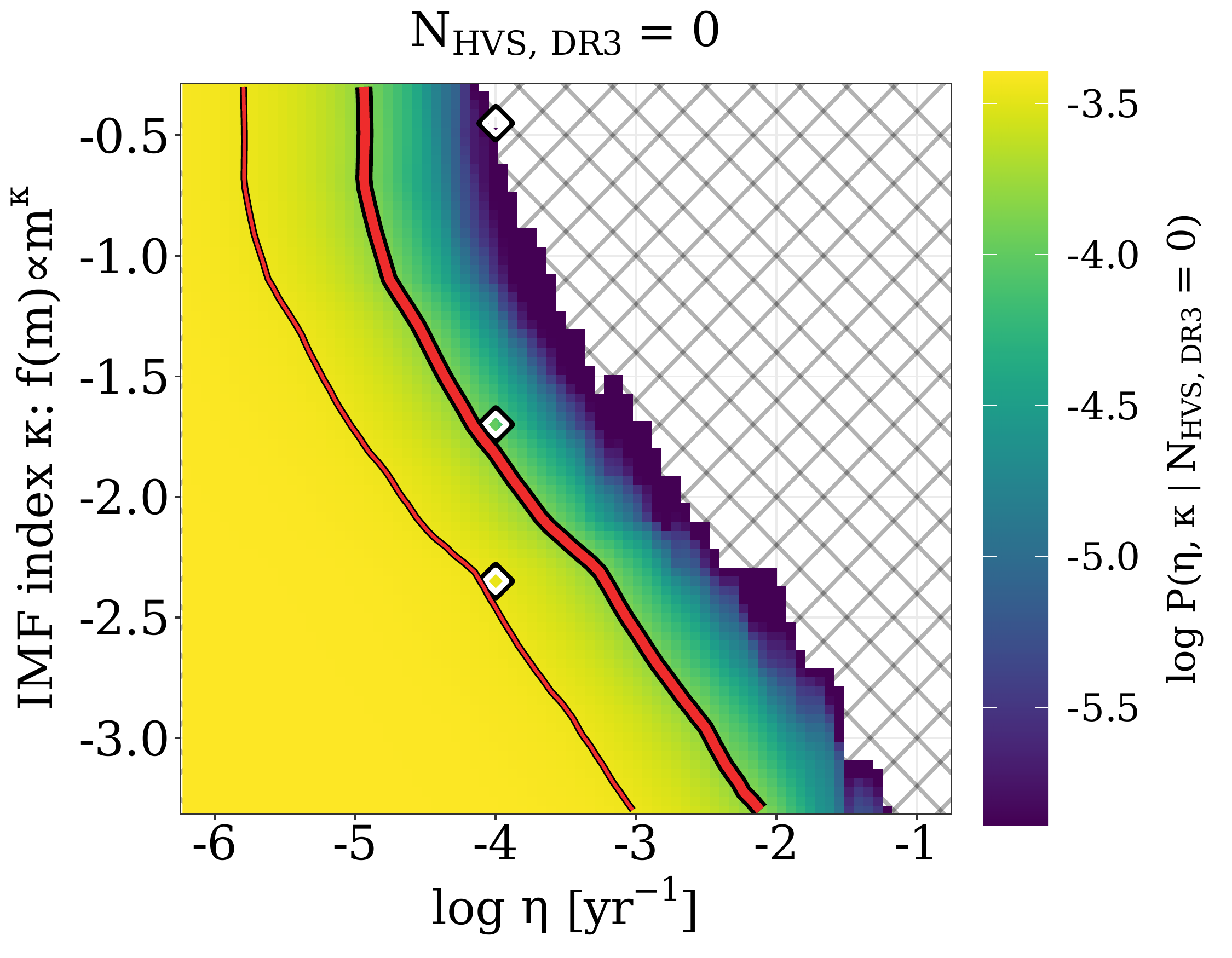}
	\caption{Joint posterior probabilities of stellar initial mass function slope $\kappa$ (for the primaries of GC binaries) and HVS ejection rate $\eta$ in light of our HVS search. The thin and thick red contours show the 68\% and 95\% Bayesian credible regions respectively. Black-and-white diamonds indicate fiducial models with $\eta=10^{-4} \, \mathrm{yr^{-1}}$ and $\kappa=-1.7$ \citep{lu+13}, $\kappa=-2.35$ \citep{salpeter55}, and $\kappa=-0.45$ \citep{bartko+10}. The hashed region shows models for which the computed posterior probability is zero.}
	\label{fig:eta_k}
\end{figure}

In \citet{Evans+22} and \citet{Evans+22b} we used the absence of unbound GC-ejected HVSs with precise parallaxes in the \textit{Gaia} DR2 and EDR3 radial velocity catalogues \citep{Marchetti+19, Marchetti21} to constrain assumptions concerning the ejection of HVSs, assuming a Hills mechanism ejection scenario. We found that the predicted number of HVSs in these surveys depends principally on the HVS ejection rate $\eta$ and on the initial mass function (IMF) of the primary star among HVS progenitor  binaries, described by a single power law with slope $\kappa$, i.e. $dN/dm \propto m^\kappa$. Constraints on these assumptions, marginalising over the other ejection parameters, can now be updated with the results of this work. Here we briefly summarize our approach to obtaining these constraints and refer to the above works for more methodological detail \citep[see also][]{Rossi+17, marchetti+18, Evans+21}.

Over a model grid in which we vary $\kappa$ and $\eta$, we adopt an MC approach; we generate HVS progenitor binaries, eject mock HVSs from the GC, compute their ages and flight times, and integrate their trajectories through the Galaxy using the same potential and integration scheme outlined in Sec. \ref{sec:results:identification}. From each star's position and astrophysical parameters, we use the MESA Isochrone and Stellar Tracks \citep[MIST;][]{dotter2016, choi+2016} to determine each mock HVS's apparent magnitudes in the {\Gaia} photometric bands. HVSs which would appear in the {\Gaia} DR3 radial velocity catalogue are selected as those brighter than $G_{\rm RVS}=14$ mag with effective temperatures spanning 3100 K $< T_{\rm eff} \leq$ 6750 K and/or those brighter than $G_{\rm RVS}=12$ mag in the effective temperature range 3100 K $< T_{\rm eff} \leq$ 14500 K \citep{Katz+22, Blomme+22}. We estimate astrometric errors using the {\Gaia} DR3 astrometric spread function of \citet{everall+2021}. To emulate the quality cuts used in our DR3 HVS search, we remove from our mock HVS population all stars which \textit{do not} satisfy $v_{\rm GC} > 700 \, \mathrm{km \ s^{-1}}$ and $\sigma_\varpi / \varpi < 0.2$. Using a Bayesian inference approach, we calculate the posterior probability for an ejection model described by ($\kappa$, $\eta$),  given the fact that zero HVSs with $v_{\rm GC} > 700 \, \mathrm{km \ s^{-1}}$ and precise parallaxes have now been unearthed in DR3. 

The results are shown in Fig. \ref{fig:eta_k}. The colorbar shows the joint posterior probability of the ejection rate $\eta$ and mass function slope $\kappa$, assuming uniform priors on each. The thin and thick red contours encompass the $68\%$ and $95\%$ credible intervals on $\eta$ and $\kappa$, respectively. The black-and-white diamonds show several fiducial models each assuming $\eta=10^{-4} \, \mathrm{yr^{-1}}$ \citep[see][]{brown15}: one with a canonical $\kappa$=-2.35 \citep{salpeter55} IMF, one with $\kappa$=-1.7 and one with $\kappa$=-0.45. The latter models follow recent estimations of the IMF shape in the inner parsec of the Galaxy from near-infrared observations; \citet{lu+13} and \citet{bartko+10}, respectively. With the non-detection of HVSs in \textit{Gaia} DR3, $\kappa$=-1.7 is strongly disfavoured for an ejection rate of $10^{-4} \, \mathrm{yr^{-1}}$, but it is still consistent with our results when the rate is $\leq$10$^{-5} \, \mathrm{yr^{-1}}$. For $\kappa$=-0.45, the upper limit on the rate is approximately $\sim 2 \times10^{-6} \, \mathrm{yr^{-1}}$. Finally, assuming the IMF of the primaries of GC binaries follows a canonical \citet{salpeter55} shape, the HVS ejection rate must be less than $8\times10^{-5} \, \mathrm{yr^{-1}}$.
Since between $50\%$ and $90\%$ of stars are in binaries, our findings for the mass distribution may be relevant for the whole GC stellar population.  Compared to the constraints obtained from the lack of HVSs in \textit{Gaia} EDR3 alone, the new results from DR3 decrease the upper limits on $\eta$ by $\simeq$0.4 dex at fixed $\kappa$.

Finally, we note that \citet{Evans+22b} show that including the positive detection of S5-HVS1 shifts the contours in $\eta$ and $\kappa$ towards the right in Fig. \ref{fig:eta_k}. While this work focuses on HVS constraints from {\Gaia} DR3 only, future works should leverage all the applicable detections (and non detections) of HVSs in Galactic surveys with well-modelled selection functions to derive joint constraints on the GC population.

\section{Discussion and Conclusions}
\label{sec:discussions}

{\Gaia} DR3 provides the largest and most homogeneous stellar spectroscopic catalogue ever produced, offering both astrometry and radial velocities for a total of 34 million sources. In this work, we derived positions and velocities in the Galactocentric rest frame for this sample of stars with the goal of discovering new unbound HVSs ejected from the GC. We summarize the main results of this work as follows:

\begin{itemize}
    
    \item We identify a clean sample of $414$ stars with accurate {\Gaia} DR3 astrometric and spectroscopic measurements, and with total velocities in the Galactocentric frame $v_\mathrm{GC} > 500$ \kms.
    
    \item By focusing on stars with large positive Galactocentric radial velocities and low Galactocentric tangential velocities, and by propagating the orbits of these stars backwards in time in the Galactic potential, we do not find any fast ($v_\mathrm{GC} > 700$ \kms) HVS candidate consistent with being ejected from the Centre of our Galaxy.
    
    \item By assuming the Hills mechanism as the sole mechanism responsible for ejecting HVSs from the GC, the non-detection of HVS candidates in this data release can be used to constrain the ejection rate $\eta$ and the power-law slope $\kappa$ of the initial mass function for the primaries of stellar binaries in the GC. Our results imply that $\eta \lesssim 8\times10^{-5}$ yr$^{-1}$ for a canonical Salpeter initial mass function. A top-heavy IMF with $\kappa=-1.7$ as suggested by \citet{lu+13} is strongly disfavoured when $\eta = 10^{-4}$ yr$^{-1}$.

\end{itemize}

Our search for HVSs focused on unbound stars, which would be easily detected due to their large values of Galactocentric radial velocities (see Fig. \ref{fig:vR_vtan}). The population of \emph{bound} HVSs \citep{bromley+06, Brown+07, kenyon+08}, i.e. stars ejected from the GC but with a total velocity that is not high enough to escape the gravitational field of the Galaxy, is expected to be dominant in {\Gaia} with respect to unbound HVSs \citep{marchetti+18, Evans+21}.These HVSs are decelerated significantly by the Galactic potential and therefore they are powerful probes to constraint its scale parameters. Unfortunately, while possible evidence for such a population could be found in the bottom-left corner of Fig. \ref{fig:E_rmin} and in the discussion in Section \ref{sec:HVSs}, a clear identification of bound HVSs is observationally challenging. These stars follow a wide variety of orbits, with multiple crossings of the Galactic midplane during their lifetime. In addition, they have Galactocentric velocity vectors pointing in all possible directions and their flight times can be of the order of several Gyrs, pushing the limits of the precision of the orbital integration (and of the accuracy of the assumed potential parameters). A successful search for bound HVSs -- outside the scope of this work -- should therefore exploit additional observations, including precise and detailed chemical abundances, which can be used in combination with kinematics to constrain the birth location of stars through chemical tagging \cite[e.g.][]{Hogg+16}, and minimize the contamination from halo stars on radial trajectories \citep{hawkins+18, Reggiani+22}.

With this work, we cannot exclude the presence of HVSs in \textit{Gaia} DR3 that are unbound and have radial velocities, but lack precise (or accurate) astrometry that can provide a robust determination of the initial conditions for orbital integration. One such example is {\Gaia} DR3 $4316462477768150144$, with $v_\mathrm{rad} = -804 \pm 7$ {\kms} and \textsc{rv\_expected\_sig\_to\_noise} $ = 14$. This star is mentioned in \citet{Katz+22} as the only star in {\Gaia} DR3 with $|v_\mathrm{vrad}| > 750$ \kms and with an expected signal-to-noise ratio above $8$. The parallax of this star is highly uncertain, $\varpi = 0.094 \pm 0.12$ mas, the astrometry is probably not accurate (\textsc{ruwe} $= 1.46$ with a $6$p solution), and spectroscopic follow-ups of this star are needed to i) estimate its distance and ii) confirm the high measured radial velocity and exclude the possibility that {\Gaia} DR3 $4316462477768150144$ is an unresolved binary. The reported {\Gaia} DR3 distance from the MARCS library is $2850^{+218}_{-537}$ pc, which, even if we trusted {\Gaia} DR3 proper motions, is not consistent with an  origin in the GC.

While this work focused exclusively on finding unbound stars coming from the GC, other astrophysical processes have been introduced to produce stars with extreme velocities without requiring a massive compact object \citep[e.g.][]{Leonard91, tauris+98, Hansen+03, abadi+09, Evans+20}. These mechanisms could explain the observed population of high velocity stars with orbits pointing away from the stellar disc \citep[e.g.][]{Irrgang+19, Marchetti21}. Focusing on the sample of 12 stars from \citet{Marchetti21} with probabilities $> 50\%$ to be unbound (when considering the zero-point correction in {\Gaia} parallaxes), we find that the total velocities from {\Gaia} DR3 are in excellent agreement with the previous ones from {\Gaia} DR2, confirming their nature as kinematic outliers. Further work is needed to explain the origin of this population.

The next {\Gaia} data release (DR4) will provide radial velocities for a sample of $\sim 150$ million stars\footnote{\url{https://www.cosmos.esa.int/web/gaia/science-performance}}, a factor $5$ more than what available now as part of {\Gaia} DR3. Combined with the more precise (and accurate) astrometry, based on observations collected during an extended baseline of 66 months\footnote{\url{https://www.cosmos.esa.int/web/gaia/release}}, it will be an ideal dataset to mine for more HVSs. Our updated model for this future {\Gaia} data release assuming a Hills ejection scenario predicts at most sixteen HVSs with radial velocity measurements,  $v_{\rm GC} > 700 \, \mathrm{km \ s^{-1}}$ and precise astrometry in {\Gaia} DR4.  Main sequence HVSs are expected to compose $\gtrsim$75 per cent of this population, with the remainder as giants \citep[c.f. fig. 7]{Evans+22b}.

This work and future ones based on the {\Gaia} stellar sample with radial velocities consider a small subset of stars in the full {\Gaia} catalogue, therefore synergy with ground-based telescopes is essential to be able to exploit the full potential of {\Gaia} data in the search and identification of HVSs. In the coming years, the advent of multi-object spectrograph facilities such as WEAVE \citep{weave}, 4MOST \citep{4MOST}, and MOONS \citep{MOONS} will provide radial velocities and stellar parameters for millions of stars in the whole sky, enabling an unprecedented knowledge of the high velocity population of stars in our Milky Way. In particular, the determination of metallicities is pivotal to gain further insights on the origin of the fastest stars, since metal-poor halo stars on radial trajectories are the main contaminants in the search for HVSs.

\section*{Acknowledgements}

The authors thank the referee, Warren R. Brown, for the careful reading of the manuscript and the useful comments. TM thanks G. Beccari, H. Boffin, N. Gentile Fusillo, T. Jerabkova, S. Janssens and the participants to the ESO {\Gaia} Coffees for useful discussions on the use of {\Gaia} data. TM thanks S. Verberne for discussions on the candidates.
TM acknowledges an ESO fellowship. 
This work has made use of data from the European Space Agency (ESA) mission
{\it Gaia} (\url{https://www.cosmos.esa.int/gaia}), processed by the {\it Gaia}
Data Processing and Analysis Consortium (DPAC,
\url{https://www.cosmos.esa.int/web/gaia/dpac/consortium}). Funding for the DPAC
has been provided by national institutions, in particular the institutions
participating in the {\it Gaia} Multilateral Agreement. EMR acknowledges that this project has received funding from the European Research Council (ERC) under the European Union’s Horizon 2020 research and innovation programme (Grant agreement No. 101002511 - VEGA P).

\emph{Software:} \textsc{numpy}\citep{numpy}, \textsc{scipy} \citep{scipy}, \textsc{Astropy} \citep{astropy}, \textsc{matplotlib} \citep{matplotlib}, \textsc{ggplot} \citep{wickham16}, \textsc{Topcat} \citep{topcat}, \textsc{gala} \citep{gala}, \textsc{galpy} \citep{bovy15}, \textsc{source\_id\_to\_orbits}.

\section*{Data Availability}

This work has made use of data from the European Space Agency (ESA) mission {\Gaia}, publicly available at the {\Gaia} archive. The simulation outputs used in this work can be shared
upon reasonable request to the corresponding author.

\bibliographystyle{mnras}
\bibliography{hvs.bib}

\appendix

\bsp

\label{lastpage}

\end{document}